\documentclass[reprint,superscriptaddress,amsmath,amssymb,aps,pre]{revtex4-2}

\usepackage{amsmath,amssymb,graphicx,enumerate,bm,dcolumn}
\usepackage{balance}
\usepackage{hyperref}
\usepackage{float}
\usepackage[english]{babel}
\usepackage{array}
\usepackage[usenames,dvipsnames]{xcolor}
\usepackage{amsfonts}

\begin{document}

\preprint{AIP/123-QED}

\title[Sample title]{Dynamics in field-induced biaxial nematic liquid crystals of board-like particles}% Force line breaks with \\

\author{Álvaro Rodríguez-Rivas}
 \email{arodriguezrivas@upo.es}
   \affiliation{ 
Department of Physical, Chemical and Natural Systems, Pablo de Olavide University, 41013 Sevilla, Spain}
\author{Alessandro Patti}%
 \email{apatti@ugr.es}
\affiliation{Department of Applied Physics, University of Granada, Fuente Nueva s/n, 18071 Granada, Spain}%
   \affiliation{Department of Chemical Engineering, The University of Manchester, Manchester M13 9PL, United Kingdom}%
\author{Alejandro Cuetos}
 \email{acuemen@upo.es}
   \affiliation{ 
Department of Physical, Chemical and Natural Systems, Pablo de Olavide University, 41013 Sevilla, Spain}

\date{\today}% It is always \today, today,

\begin{abstract}

Biaxial nematic ($\rm N_B$) liquid crystals have been indicated as promising candidates for the design of next-generation displays with novel electro-optical properties and faster switching times. While at the molecular scale their existence is still under debate, experimental evidence, supported by theory and simulation, has unambiguously proved that suitable colloidal particles can indeed form $\rm N_B$ fluids under specific conditions. While this discovery has sparked a widespread interest in the characterisation of the phase behaviour of $\rm N_B$ liquid crystals, significantly less attention has been devoted to the study of their transport properties. To bridge this gap, by Dynamic Monte Carlo simulations we have investigated the equilibrium dynamics of field-induced $\rm N_B$ phases comprising monodisperse hard cuboids. 
In particular, we calculated the long-time self-diffusion coefficients of cuboids over a wide range of anisotropies, spanning prolate to oblate geometries. Additionally, we have compared these diffusivities with those that, upon switching the external field off, are measured in the thermodynamically-stable isotropic or uniaxial nematic phases at the same density. Our results indicate that while prolate cuboids diffuse significantly faster in biaxial nematics than in less ordered fluids, we do not observe such an increase with oblate cuboids at high packing fractions. We show that these changes are most likely due to the field-induced freezing of the axes perpendicular to the nematic director, along with a substantial increase in the ordering of the resulting $\rm N_B$ phase.
\end{abstract}

                            
\maketitle

%%%%%%%%%%%%%%%%%%%%%%%%%%%%%%%%%%%%%
%%%%%%%%%% INTRODUCTION %%%%%%%%%%%%%
%%%%%%%%%%%%%%%%%%%%%%%%%%%%%%%%%%%%%

\section{\label{sec:intro}Introduction}
Colloids are two-phase systems comprising a phase homogeneously dispersed throughout a continuous medium. The dispersed phase can be observed in the form of droplets, particles or bubbles depending on whether it is a liquid, solid or gas, respectively. Similarly, the dispersing phase can also exist as a fluid or a solid. The particular case of solid particles dispersed in a liquid is referred to as colloidal sol or simply sol. This family of colloids finds broad application in the design of numerous industrially relevant formulations, including paints, foods, pharmaceuticals and personal-care products. Especially fascinating is the case of sols comprising anisotropic particles for they can form long-range ordered mesophases, referred to as liquid crystals (LCs). In particular, nematic LCs exhibit a merely orientational ordering, with all particles almost completely aligned along a common direction, but randomly distributed in the dispersing fluid \cite{onsager1949, degennes1974}.

In order to fully control their properties, it is important to understand how sols behave under equilibrium and out-of-equilibrium conditions. More specifically, one should know their phase behaviour and how this can be perturbed by external stimuli such as a temperature gradient, a shear, a gravitational or an electromagnetic field. External stimuli can be as weak as a few $k_BT$ per particle, with $k_B \approx 1.381 \times 10^{-23}$ JK$^{-1}$ the Boltzmann's constant and $T$ the absolute temperature. These apparently tiny amounts of energy are sufficient to spark dramatic changes affecting the organisation of the dispersed particles in the fluid phase and to eventually lead to ordered-disordered phase transitions. Considering that, in most practical applications, colloidal sols are not at the thermodynamic equilibrium, calculating their phase diagrams is indeed a necessary step to ponder their use in formulation technology, but it is far from being sufficient. It is therefore crucial to investigate how colloidal sols respond to external forces and how their dynamical, structural and rheological properties change as a result of a given perturbation. This is especially important in sols comprising anisotropic (non-spherical) particles, because the application of an external stimulus, including confinement, can order them along a preferential direction, eventually manipulating the system's ordering and the complete spectrum of its properties \cite{teun2013, vutukuri2014, odriozola2020, teixeira2021}. For instance, upon application of an external shear, isotropic suspensions of rod-like or disk-like particles can be transformed into nematic or positionally ordered LCs, such as smectic or columnar LCs, respectively \cite{ripoll2008}. The more complex the particle geometry is, the less obvious the system's response to an external stimulus will be. In particular, under the action of an external field, uniaxial particles (\textit{e.g.} colloidal needles)  can orient along their major axis and thus form prolate nematic LCs, whereas biaxial particles (\textit{e.g.} nanoboards) can either orient along their major or minor axis and form prolate or oblate nematic LCs \cite{lettinga2005, baza2020, parisi2021}. If both sets of axes are oriented, these systems are referred to as biaxial nematic ($\rm N_B$) LCs. 

The earliest studies of $\rm N_B$ phases date back to 1970, when Freiser theoretically predicted their existence by generalising the Maier-Saupe theory to incorporate the effect of molecular biaxiality on phase behaviour \cite{Freiser1970}. More than fifty years later, the interest in this family of LCs is still vivid, especially because an unambiguous evidence of the existence of $\rm N_B$ phases in thermotropic systems is still pending \cite{jakli2018}. The pioneering work by Freiser was then followed by other equally elegant theories that indeed postulated the thermodynamic stability of $\rm N_B$ LCs as well as the possible existence of a direct $\rm I$-to-$\rm N_B$ transition for self-dual particles, whose geometry is exactly in between oblate and prolate \cite{Straley1974, Mulder1989, Taylor1991}. It should be noticed that these theories were developed either assuming the restricted-orientation (Zwanzig) model, which only allows six orthogonal particle orientations \cite{Zwanzig1963}, or neglecting the existence of positionally ordered phases, such as smectic LCs and crystals. Following the first unambiguous experimental evidence of the existence of biaxial nematics in polydisperse systems of board-like colloidal particles \cite{vroege2009}, more recent theories investigated the effect of size dispersity on the stability of the $\rm N_B$ phase, but still restricting particle orientation \cite{Vanakaras2003, Belli2011, Gonzalez-Pinto2015}. Monte Carlo (MC) simulations of freely-rotating cuboids finally showed that only by introducing a significant degree of size dispersity \cite{effran2020} could biaxial nematics be observed, confirming former experimental observations \cite{vroege2009} and theoretical intuitions \cite{Belli2011}. Simulations also showed that monodisperse or bi-disperse suspensions of freely-rotating board-like particles cannot form $\rm N_B$ phases \cite{cuetos2017, patti2018}, unless particle anisotropy is extremely large \cite{dussi2018}. It should be anyway noticed that experiments on highly uniform colloidal cuboids of extreme anisotropy, probing stacking rather than bulk behaviour, did not find evidence of the existence of biaxial nematics \cite{nie2018}. Alternatively, an external field imposing alignment of one of the three particle axes can transform isotropic or uniaxial nematic phases into biaxial nematics \cite{cuetos2019}. 

While a very significant interest has been devoted to the analysis of their phase behaviour, the study of the dynamics of cuboids in LC phases has received considerably less attention. In this work, we study the dynamics of a family of colloidal cuboids that form $\rm N_B$ LCs under the application of an external field. To the best of our knowledge, transport properties in biaxial nematic phases have not been studied in the past. In particular, we characterise the resulting mobility of oblate, prolate and self-dual-shaped cuboids in the direction of the field applied and perpendicularly to it. To this end, we employ the Dynamic Monte Carlo (DMC) simulation method, a stochastic technique that can qualitatively and quantitatively reproduce the Brownian dynamics of colloids under well-specified elementary rotational and translational moves. We originally developed the DMC technique for investigating the dynamics of monodisperse \cite{patti2012} and polydisperse \cite{cuetos2015} colloidal sols at equilibrium and then extended it to the study of unsteady-state processes \cite{corbett2018}, heterogeneous systems \cite{garciadaza2020} and microrheology \cite{garciadaza2022}. In its final form, DMC can basically be applied to assess the dynamics of any colloidal suspensions of hard or soft particles. We have already applied it to study the dynamics of cuboids in the bulk and under confinement \cite{cuetos2020, patti2021, tonti2021} as well as the uniaxial-to-biaxial switching upon application of an external field \cite{effran2021}. However, the equilibrium dynamics of cuboids forming field-stabilised $\rm N_B$ LCs has not yet been investigated. Our intention is to
bridge this gap in the present paper, which is organised as follows. In Section~\ref{sec:model}, we introduce the model and simulation methods applied to equilibrate the systems of interest and investigate their dynamics. Because the DMC technique has been presented elsewhere, here we only remind the key results that are strictly necessary to follow our arguments and remind the interested reader to our previous works for details. In Section~\ref{sec:results}, we characterise the dynamics by estimating the ability of particles to diffuse at long times as a function of their geometry.  Finally, in Section~\ref{sec:conclusions} we draw our conclusions.

%%%%%%%%%%%%%%%%%%%%%%%%%%%%%%%%%%%%%
%%%%%%%%%% MODEL AND SIMS %%%%%%%%%%%
%%%%%%%%%%%%%%%%%%%%%%%%%%%%%%%%%%%%%

\section{\label{sec:model}Model and simulation details}

\begin{figure}[h!]
\includegraphics[width=.90 \columnwidth]{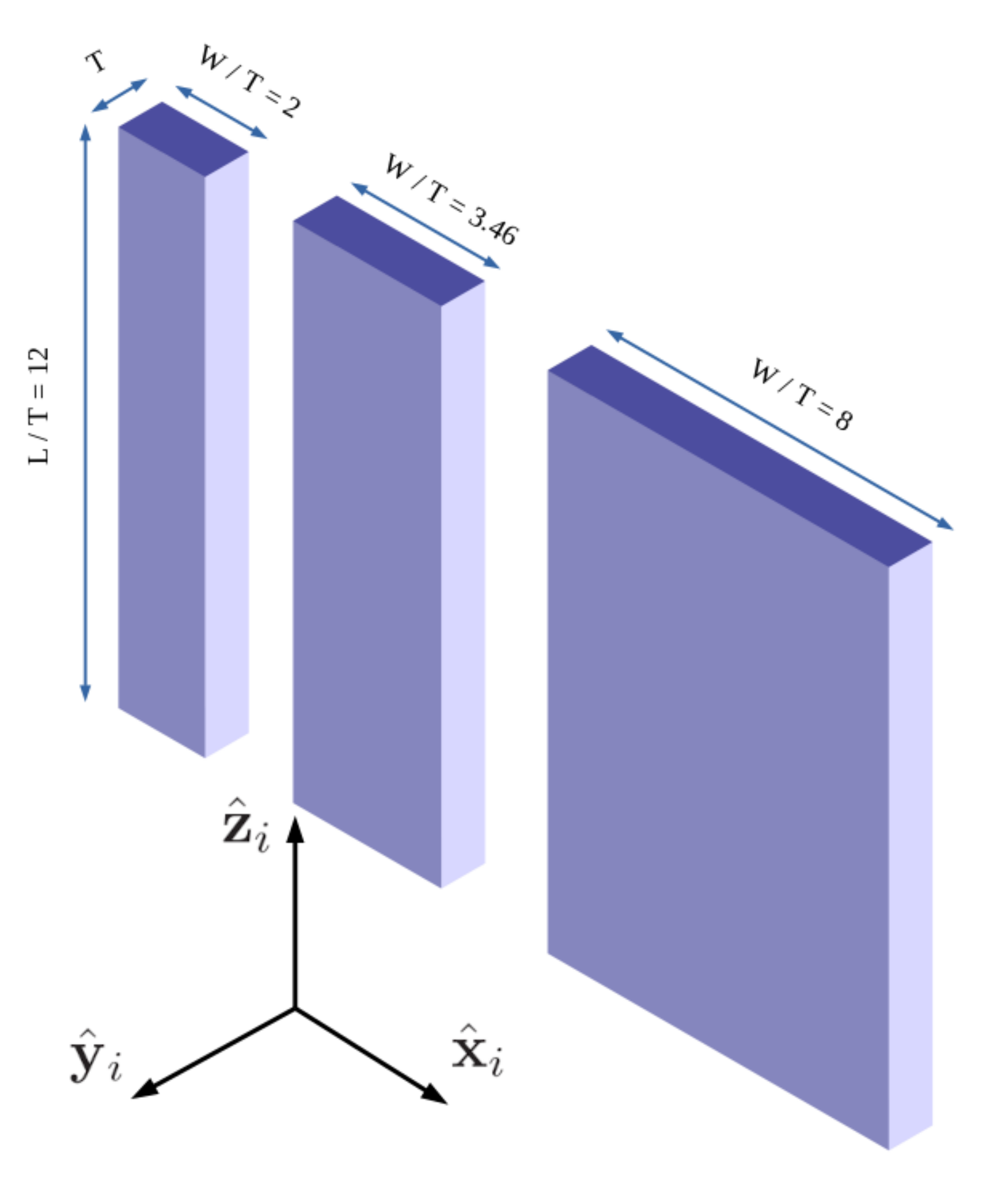}
\caption{Model HBPs with thickness $T$, length $L/T = 12$ and
width $W/T=2, 3.46$ and $8$. The unit vectors $\hat{\bf{x}}_i$, $\hat{\bf{y}}_i$, and $\hat{\bf{z}}_i$ indicate the orientation of $W$, $T$, and $L$, respectively.}\label{fig:CUBOIDE} 
\end{figure}

In this work, we study the equilibrium dynamics of $\rm N_B$ LCs of monodisperse colloidal cuboids. More specifically, we are interested in characterising the dynamical properties of biaxial nematic fluids induced by applying an external field to isotropic (I) or uniaxial nematic ($\rm N_U$) phases that would spontaneously form if the field was absent. To this end, the cuboids have been modelled as hard board-like particles (HBPs) of aspect ratio $L^{*} \equiv L/T =12$, where $L$ is the particle length and $T$ the particle thickness and system unit length. To study the impact of geometry on the resulting dynamics, the reduced width, $W^* \equiv W/T$, is varied between 2 and 8 (see Fig.\,\ref{fig:CUBOIDE} for details). In particular, self-dual shaped particles, with $W^*=\sqrt{L^*} \approx 3.46$, are exactly at the crossover between prolate ($W^*<\sqrt{L^*}$) and oblate ($W^*>\sqrt{L^*}$) particles.

Because particles interact via a hard potential, their phase behaviour \cite{cuetos2017} is fully determined by shape anisotropy and packing fraction $\eta \equiv \nu_{0}N_{p}/V$, with $\nu_{0}=LWT$ the particle volume, $V$ the simulation box volume and $N_{p}$ the number of HBPs, which ranges between $1152$ and $4608$ depending on $W^*$. The packing fraction has been set according to the system phase diagram, available to the interested reader in Ref.\,\cite{cuetos2017}. Specifically, for $\rm N_B$ phases obtained by field-induced reorientation of $\rm N_U$ phases, we set $\eta=0.340$ for the complete spectrum of geometries studied. We note that this value of the packing fraction is the same as that recently used to investigate the equilibrium dynamics of thermodynamically stable $\rm N_U$ LCs \cite{cuetos2020} and it is hence especially appropriate to ponder the effect of orientational ordering (biaxiality \textit{vs} uniaxiality) on long-time particle dynamics. By contrast, the dynamics of $\rm N_B$ phases obtained by applying an external field to I phases has been studied at different packing fractions, between 0.220 and 0.307, depending on $W^*$,  corresponding to state points that are just below the I-$\rm N_U$ transition \cite{cuetos2017}. In the following, the abbreviations $\rm N^U_B$ and $\rm N_B^I$ will be employed to indicate field-induced $\rm N_B$ LCs obtained from $\rm N_U$ and I phases, respectively.

% CUBOIDE

To induce the onset of the $\rm N_B$ phase, we applied an external field that promotes the alignment of the particle intermediate axis $\hat{\bf{x}}_i$ along the field direction $\hat{\bf{e}}$ \cite{cuetos2019}:

\begin{equation}
U_{\rm ext}=\frac{\varepsilon_{f}}{2}\left(1-3\cdot (\hat{\bf{x}}_{i}\cdot\hat{\bf{e}})^{2}\right),
\label{Uext1}
\end{equation}

\noindent where $\varepsilon_{f}$ indicates the field strength. In order to ensure the collective rearrangement of the fluid and the stabilisation of well-defined biaxial nematics in all the cases studied in this work, we have set $\beta\varepsilon_{f}=2$, with $\beta^{-1}$ the energy unit. Weaker field intensities, with $\beta \varepsilon_{f} \le 1 $ generally give weakly ordered $\rm N_B$ phases regardless the particle width \cite{cuetos2019}. 

The simulations of all systems consisted of an equilibration run followed by a production run. Thus, we first equilibrated I and $\rm N_U$ phases with the field switched off, then we switched the field on to induce biaxiality, and finally produced the time trajectory of the so-obtained $\rm N_B$ fluids. To equilibrate the systems, with the field on or off, we performed standard MC simulations of typically $10^5$ cycles, with one cycle consisting of $N_p$ independent attempts to displace the particle center of mass and/or reorient its axes. Shall an attempted move lead to an overlap between two particles, then the move is rejected. If this is not the case, the move is accepted according to the Metropolis algorithm \cite{metropolis1953, FRENKEL200223}, which incorporates the energy difference between new and old configurations as determined by the external field defined in Eq.\,\ref{Uext1}. No other energy contributions are considered due to the hard-core nature of the particles. The interested reader is referred to Refs.\,\cite{GLM96} and \cite{JE05} for details on the estimation of overlaps between cuboids. All simulations were run in the canonical ensemble and in orthogonal boxes with periodic boundaries. To assess equilibration, we monitored the stabilisation of the packing fraction and order parameters, which are obtained from diagonalisation of the following second-rank symmetric tensor\cite{EF84}:

\begin{equation}
{\bf{Q}^{\lambda\lambda}}=\frac{1}{2N_{p}}\left<\sum_{i=1}^{N_{p}}(3\hat{\lambda}_{i}\cdot\hat{\lambda}_{i}-\bf{I})\right>,
\label{eq:tensor}
\end{equation}  

\noindent where $\hat{\lambda}_{i}=\hat{\bf{x}}_i, \hat{\bf{y}}_i, \hat{\bf{z}}_i$ and $\bf{I}$ is the identity tensor. The resulting eigenvalues $S_{2,W}$, $S_{2,T}$ and $S_{2,L}$ identify the uniaxial order parameters associated to the collective orientation of each particle axes, while the corresponding eigenvectors $\hat{\bf{m}}$, $\hat{\bf{p}}$ and $\hat{\bf{n}}$ represent the respective nematic directors. A relatively large value of at least one of these order parameters is the evidence of the alignment of the corresponding particle axis along the direction defined by the associate nematic director. Similarly, the biaxial order parameters $B_{2,W}$, $B_{2,T}$ and $B_{2,L}$ identify the occurrence of biaxiality by measuring the fluctuations of the particle axes perpendicular to the nematic director associated to the corresponding eigenvalue of the tensor defined in Eq.\,\ref{eq:tensor}. For example, if HBPs aligned along their $\hat{\bf{z}}$ axis, then $S_{2,L}$ would be the largest uniaxial order parameter and $\hat{\bf{n}}$ the main nematic director. In this case, the biaxial character of the system can be determined as $B_{2,L}=(\hat{\bf{m}}\cdot\hat{\bf{Q}}^{xx}\cdot\hat{\bf{m}} + \hat{\bf{p}}\cdot\hat{\bf{Q}}^{yy}\cdot\hat{\bf{p}} - \hat{\bf{m}}\cdot\hat{\bf{Q}}^{yy}\cdot\hat{\bf{m}} -\hat{\bf{p}}\cdot\hat{\bf{Q}}^{xx}\cdot\hat{\bf{p}})/3$. Similar expressions can be used to obtain $B_{2,W}$ and $B_{2,T}$. However, to assess the onset of biaxiality is not necessary to monitor the three biaxial order parameters, but only that associated to the axis displaying the largest uniaxial order parameter \cite{allen1990, camp1999, teixeira2006}. The values of uniaxial and biaxial order parameters of the $\rm N_B$ phases explored in this work are consistent with those obtained 

in previous works\cite{cuetos2017, patti2018, cuetos2019}. 

Following equilibration, configurations of $\rm N_B^I$ and $\rm N^U_B$ fluids have been employed as starting points for the production of time trajectories and the estimation of the dynamical properties of interest. Similarly to MC simulations, each DMC cycle consists of $N_p$ independent attempts to move randomly-selected HBPs. Nevertheless, in this case, translational and rotational moves are always attempted simultaneously. We showed that this choice satisfies the simple balance condition, which is a sufficient and necessary condition \cite{Manousiouthakis1999}, and does not alter the Boltzmann distribution of the ensemble \cite{patti2012}.

%TABLA II
\begin{table*}[ht!]
\caption{Table of infinite-dilution translational and orientational diffusion coefficients of HBPs as obtained from HYDRO$^{++}$ \cite{CT99,GTREO07}. Particle reduced length is $L^*=12$, $\tau=T^{3}\beta\mu$ is the time unit and $\mu$ the viscosity of the solvent.}\label{tab:table1} 
%\begin{ruledtabular}
\begin{tabular}{lcccccc}
$W^{*}$ & $D^{tra}_{T}$ & $D^{tra}_{W}$ & $D^{tra}_{L}$ & $D^{rot}_{T}$ & $D^{rot}_{W}$ & $D^{rot}_{L}$\\
$ $ & $(10^{-2}T^2/\tau)$ & $(10^{-2}T^2/\tau)$ & $(10^{-2}T^2/\tau)$ & $(10^{-4}/\tau)$ & $(10^{-4}/\tau)$ & $(10^{-3}/\tau)$\\
\hline
2 & $1.79$ & $1.95$ & $2.54$ & $8.63$ & $7.87$ & $8.43$\\
2.5 & $1.67$ & $1.88$ & $2.36$ & $7.90$ & $7.08$ & $5.78$\\
3 & $1.57$ & $1.81$ & $2.22$ & $7.26$ & $6.45$ & $4.18$\\
3.46 & $1.40$ & $1.80$ & $2.20$ & $6.70$ & $6.00$ & $3.20$\\
4 & $1.40$ & $1.71$ & $1.99$ & $6.19$ & $5.54$ & $2.45$\\
6 & $1.15$ & $1.52$ & $1.67$ & $4.60$ & $4.34$ & $1.11$\\
8 & $0.94$ & $1.38$ & $1.46$ & $3.49$ & $3.57$ & $0.63$\\
\end{tabular}

%\end{ruledtabular}
\end{table*}

In particular, elementary displacements and rotations are generated from uniform distributions that depend on the translational and rotational diffusion coefficients at infinite dilution, $D^{\rm tra}_{\alpha}$ and $D^{\rm rot}_{\alpha}$ respectively, with $\alpha=L,W,T$. For translations, the elementary displacement is defined by $\delta {\bf r} = X_{W} \hat{\bf{x}} + X_{T} \hat{\bf{y}} + X_{L}\hat{\bf{z}}$, decoupling into the three unitary directions  and restricted by the maximum displacements $|X_{\alpha}| \le \sqrt{2D^{\rm tra}_{\alpha} \delta t_{MC}}$. In the case of rotations, the particles axes are reoriented by three consecutive rigid rotations around $L$, $W$ and $T$, respectively, with the maximum rotation around each particle axis $|Y_{\alpha}| \le \sqrt{2D^{\rm rot}_{\alpha}\delta t_{MC}}$. In all the cases, $\delta t_{\rm MC} = 10^{-2} \tau$, with $\tau=T^3\beta\mu$ the time unit and $\mu$ the viscosity of the solvent. The infinite-dilution diffusion coefficients have been obtained by using the open-source software HYDRO$^{++}$\cite{CT99,GTREO07}, and are shown in table \protect\ref{tab:table1}. Finally, to express the results as a function of an actual Brownian dynamics timescale, the MC timescale need to be rescaled, using the acceptance rate $\mathcal{A}$, as \cite{patti2012}:

\begin{equation}
\delta t_{BD}= \frac{\mathcal{A}}{3}\delta t_{MC}
\label{eq:acc}
\end{equation}

In this study, we have calculated a set of dynamical observables. These include the isotropic mean-squared displacement (MSD) as well as its parallel and perpendicular components with respect to the nematic director $\hat{\bf{m}}$, $\hat{\bf{p}}$ or $\hat\bf{n}$.

\begin{eqnarray}\label{eq-msd}
\displaystyle{\left<\Delta r^{2}\left(t\right)\right>} \,=\, \frac{1}{N_p}\displaystyle{\sum_{i=1}^{N_{p}}\left<\left|\bf{r}_{i}\left(t\right)-\bf{r}_{i}\left(0\right)\right|^{2}\right>},\\
\displaystyle{\left<\Delta r^{2}_{\parallel}\left(t\right)\right>} \,=\, \frac{1}{N_p}\displaystyle{\sum_{i=1}^{N_{p}}\left<\left|\bf{r}_{\parallel,i}\left(t\right)-\bf{r}_{\parallel,i}\left(0\right)\right|^{2}\right>},\\
\displaystyle{\left<\Delta r^{2}_{\perp}\left(t\right)\right>} \,=\, \frac{1}{2N_p}\displaystyle{\sum_{i=1}^{N_{p}}\left<\left|\bf{r}_{\perp,i}\left(t\right)-\bf{r}_{\perp,i}\left(0\right)\right|^{2}\right>}
\end{eqnarray}

\noindent where $\left< \dots \right>$ denotes average over 200 independent trajectories, while $\bf{r}_{\parallel,i}$ and $\bf{r}_{\perp,i}$ are, respectively, the projections of the displacement of particle $i$ in the directions parallel and perpendicular to a given nematic director. These directional MSDs are especially useful if the system exhibits nematic ordering as they provide an insight into the particle translational self-diffusion coefficients, which are proportional to the long-time slope of the MSD $vs$ time, that is $D=\lim_{t\to +\infty}\left<\Delta r^{2}\right>/2dt$, where $d=1$ for parallel and perpendicular MSDs, and $d=3$ for the total MSD.

We have also calculated the orientational diffusion coefficients which provides information on the particle orientational relaxation. Due to their biaxial geometry, HBPs exhibit three independent orientational diffusion coefficients, corresponding to the re-orientation of the three particle axes $\hat{\bf{x}}$, $\hat{\bf{y}}$ and $\hat{\bf{z}}$. These coefficients have been calculated via the orientational time-correlation functions \cite{HM06,H19}

\begin{equation}
C_{\alpha}=\langle P_{1}\left[\hat{\bf{e}}_{\alpha}(t)\cdot \hat{\bf{e}}_{\alpha}(0)\right]\rangle
\label{eq:c}
\end{equation}

\noindent where $\hat{\bf{e}}_{\alpha}=\hat{\bf{x}}_i$, $\hat{\bf{y}}_i$ or $\hat{\bf{z}}_i$, $P_{1}$ is the first Legendre polynomial, and the brackets indicate ensemble averages over $N_p$ particles and $200$ different trajectories. For each particle axes, the corresponding relaxation time has been calculated as follows:

\begin{equation}
\tau_{\alpha}=\int^{\infty}_{0} C_{\alpha}dt
\label{eq:tau}
\end{equation}

\noindent and the three different long time orientational diffusion coefficients are given by $D^{or}_{\alpha}=1/2\tau_{\alpha}$. These diffusion coefficients indicate how fast the reorientation of each of the particle axes is at long times. By contrast, the rotation coefficients $D^{\rm rot}_{\alpha}$ are related to the rigid rotation of the cuboidal particle around the axis $\alpha= L, W$ or $T$.

%%%%%%%%%%%%%%%%%%%%%%%%%%%%%%%%%%%%
%%%%%%%%%%%%%%%%%%%%%%%%%%%%%%%%%%%%
%%%%%%%%%%%%%%%%%%%%%%%%%%%%%%%%%%%%

\section{\label{sec:results}Results}

\begin{figure}[h!]
\includegraphics[width=1.00 \columnwidth]{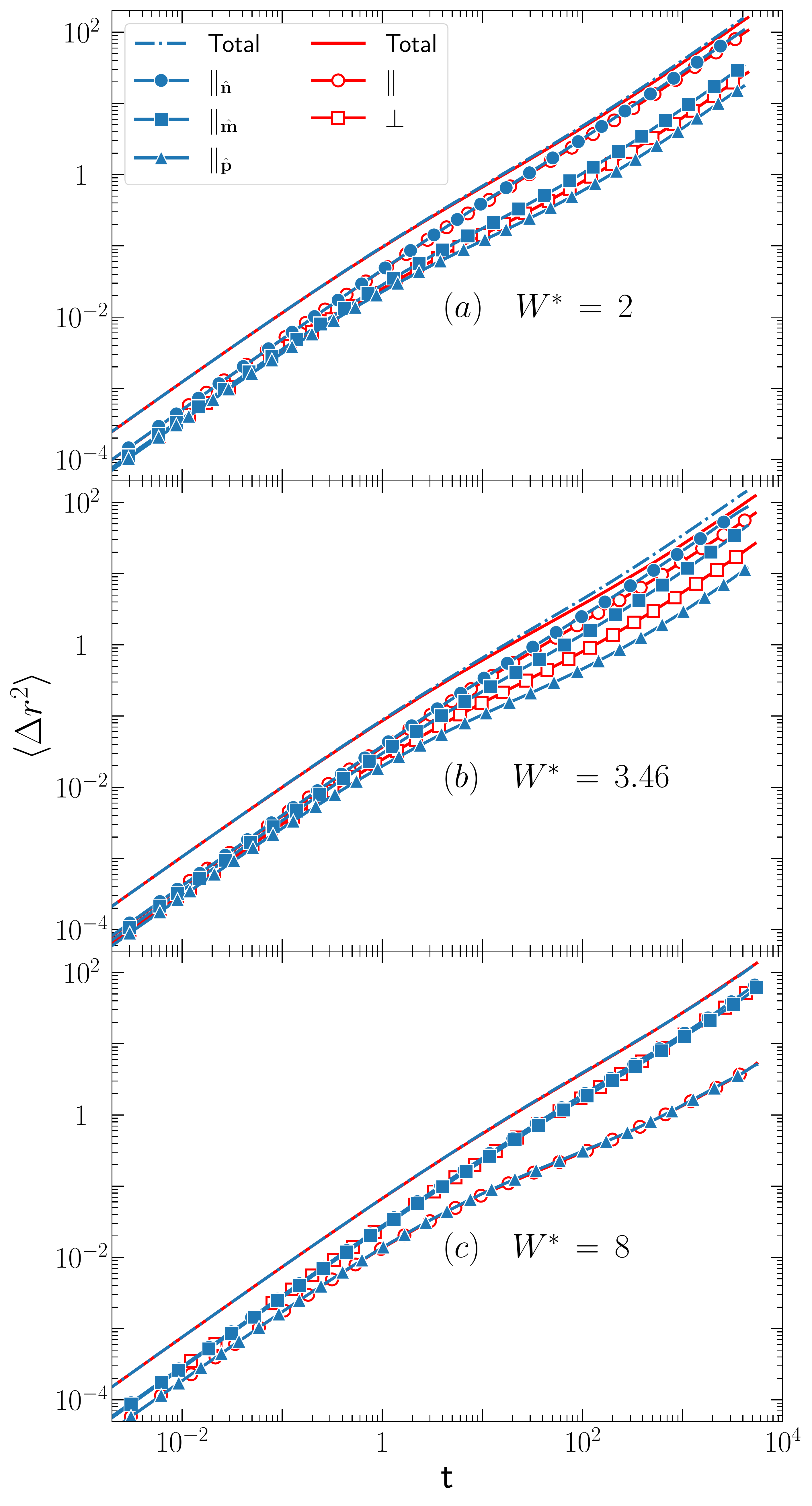}
\caption{MSDs of HBPs in $\rm N_U$ ($\epsilon_{f}^{*} = 0$) and field-induced $\rm N_B^U$ ($\epsilon_{f}^{*} = 2$) phases, both at $\eta = 0.34$. Empty circles and squares correspond, respectively, to the parallel and perpendicular MSD in the $\rm N_{U}$ phase. Solid circles, squares and triangles refer to the MSDs obtained in the $\rm N_B^U$ phase along the nematic directors $\hat{\bf n}$, $\hat{\bf m}$ and $\hat{\bf p}$, respectively. Isotropic MSDs are represented by solid ($\rm N_{U}$) and dotted-dashed ($\rm N_B^U$) lines.}
\label{fig:MSDNEM}
\end{figure}

In the present section, our goal is understanding to what extent the dynamics of HBPs in biaxial nematics exhibits distinctive details that are not detected in less ordered fluids. In addition, we would like to ascertain the dependence of particle mobility on shape anisotropy and therefore the existence of especially suitable geometries, among those investigated here,  that favour rotational and translational diffusion as compared to others. To this end, we have determined the MSD of a wide spectrum of shapes, spanning rod-like to disk-like particles, by tuning the particle width and keeping constant thickness and length. In Fig.\,\ref{fig:MSDNEM}, we compare the MSDs obtained in the $\rm N_U$ phase at $\eta=0.34$ with those in the $\rm N_B^U$ phase at the same packing fraction. At this $\eta$ value the nematic phase is stable over the whole range of $W/T$ studied \cite{cuetos2017}. The former have been calculated in thermodynamically stable phases with no field applied ($\epsilon_f^* \equiv \beta\epsilon_f=0$), whereas the latter have been obtained upon application of the external field $U_{\rm ext}$, with intensity $\epsilon_f^*=2$. In particular, three different particle anisotropies are shown in this figure: prolate HBPs with $W^*=2$ (top frame), self-dual shaped HBPs with $W^*=\sqrt{L^*}\approx 3.46$ (middle frame) and oblate HBPs with $W^*=8$ (bottom frame). The $\rm N_U$ phase at $\eta=0.34$ was shown to exhibit a prolate character for $W^* \leq \sqrt{L^*}$ \cite{cuetos2017}, with the particle unit vectors $\hat{\bf{z}}_i$ strongly correlated along the nematic director $\hat{\bf{n}}$, but $\hat{\bf{x}}_i$ and $\hat{\bf{y}}_i$ randomly oriented. By contrast, for $W^* > \sqrt{L^*}$, the $\rm N_U$ phase exhibits a clearly oblate character, with the particle unit vectors $\hat{\bf{y}}_i$ strongly correlated along the nematic director $\hat{\bf{p}}$, while $\hat{\bf{x}}_i$ and $\hat{\bf{z}}_i$ almost completely uncorrelated. For simplicity, prolate and oblate uniaxial nematic LCs are respectively indicated as $\rm N_U^+$ and $\rm N_U^-$. Consequently, it makes sense to calculate the MSD along $\hat{\bf{n}}$ in $\rm N_U^+$ phases or $\hat{\bf{p}}$ in $\rm N_U^-$ phases as well as in directions perpendicular to these nematic directors. Parallel and perpendicular MSDs in these uniaxial phases are reported in the three frames of Fig.\,\ref{fig:MSDNEM} and indicated by empty circles and squares, respectively. In agreement with our recent works on hard cuboids \cite{cuetos2020} and soft repulsive spherocylinders \cite{morillo2019}, the tendencies of Fig.\,\ref{fig:MSDNEM} confirm that, in $\rm N_U^+$ phases (frames (a) and (b)), the long-time particle mobility along $\hat{\bf{n}}$ is more pronounced than that in planes perpendicular to $\hat{\bf{n}}$. By contrast, the MSDs obtained in $\rm N_U^-$ phases (frame (c)) indicate that oblate HBPs prefer to move in planes perpendicular to the relevant nematic director rather than parallel to it.

Upon application of $U_{\rm ext}$, the system experiences a re-equilibration, with the particles forced to reorient their unit vector $\hat{\bf{x}}$ along the direction of the field. Such a transitory out-of-equilibrium condition is then followed by a new equilibrium state where biaxiality is observed if the field intensity is sufficiently strong \cite{cuetos2019}. In the so-obtained $\rm N_B^U$ phase, the symmetry in the particle orientation is broken and, consequently, it makes sense to study the dynamics along three mutually perpendicular nematic directors, $\hat{\bf{m}}$, $\hat{\bf{p}}$ and $\hat{\bf{n}}$, with the unit vectors $\hat{\bf{x}}$, $\hat{\bf{y}}$ and $\hat{\bf{z}}$ preferentially oriented, respectively, along each of them. Due to this symmetry breaking, particle diffusion is not expected to be isotropic, but instead to change along the directions defined by the three nematic directors. To test this hypothesis, we have investigated the equilibrium dynamics of the field-induced $\rm N_B^U$ phases and the resulting MSDs are reported in Fig.\,\ref{fig:MSDNEM} for prolate, self-dual shaped and oblate HBPs. In particular, in each frame we show the MSD parallel to $\hat{\bf{m}}$ (solid squares), $\hat{\bf{p}}$ (solid triangles) and $\hat{\bf{n}}$ (solid circles). In the top frame, where we analyse the dynamics of rod-like HBPs, the MSD parallel to $\hat{\bf{n}}$ does not seem to be especially affected by the presence of the external field as it increases very slightly, at long times, as compared to the MSD calculated in the parental $\rm N_U^+$ phase. Similar tendencies are also noticed in systems of self-dual shaped HBPs, although here the difference between the two parallel MSDs is more significant, and in systems of oblate HBPs, where the  dynamics along $\hat{\bf{p}}$ in $\rm N_U^-$ and $\rm N_B^U$ phases are practically indistinguishable. 

To fully appreciate the effect of the field-induced phase biaxiality on the dynamics of HBPs, we now analyse the MSDs along the directions perpendicular to the main nematic director. While in the $\rm N_U^+$ and $\rm N_U^-$ phases all these directions are equivalent, in the $\rm N_B^U$ phase there are two preferential directions, which correspond to the nematic directors, $\hat{\bf{p}}$ and $\hat{\bf{m}}$ in case of prolate nematics or $\hat{\bf{m}}$ and $\hat{\bf{n}}$ for oblate nematics. The resulting MSDs along these directors are strongly determined by the intensity of the applied field, which is always coupled to the particle unit vector $\hat{\bf{x}}_i$ and thus aligned with the nematic director $\hat{\bf{m}}$ in $\rm N_B^U$ phases of prolate and oblate HBPs. By imposing reorientation of the unit vectors $\hat{\bf{x}}_i$, the field is also forcing the reorientation of the unit vectors $\hat{\bf{y}}_i$ (prolate case) or $\hat{\bf{z}}_i$ (oblate case), thus intimately correlating the dynamics of particles along these two directions with their geometry. It follows that particle anisotropy contributes to determine the effect of the applied field on the dynamics and the resulting differences observed in the directional MSDs should be assessed with this in mind. If we analyse the dynamics along the directors that are directly affected by the external field, we observe that, in $\rm N_B^U$ phases of prolate HBPs, the long-time MSD in the direction of $\hat{\bf{m}}$ (and hence of $\hat{\bf{e}}$) is larger than that in the direction of $\hat{\bf{p}}$. This difference decreases from $W^*=3.46$ to $W^*=2$ and would most likely disappear at $W^*=1$ with the rod-like cuboids exhibiting a squared cross section. By contrast, in $\rm N_B^U$ phases of oblate HBPs ($W^*=8$), with the field forcing the reorientation of $\hat{\bf{m}}$ and $\hat{\bf{n}}$, the MSDs parallel to these two directions are very similar to each other and to the corresponding MSD in the parental $\rm N_U^-$ phase. Finally, if we compare the mobility along the relevant nematic directors of the initial uniaxial phases with that of the resulting biaxial phases, we notice that the long-time MSD along $\hat{\bf{n}}$ of prolate and self-dual shaped HBPs is larger in $\rm N_B^U$ than in $\rm N_U^+$ phases, but no difference is detected between the long-time MSDs along $\hat{\bf{p}}$ measured in $\rm N_B^U$ and $\rm N_U^-$ phases of oblate HBPs. In other words, applying an external field does not have any tangible impact on the dynamics of oblate HBPs, which exhibit essentially the same MSDs in uniaxial and biaxial nematics. This behaviour has also been observed at $W^*=4$ and $6$. 

\begin{figure}[h!]
   \includegraphics[width=1.00 \columnwidth]{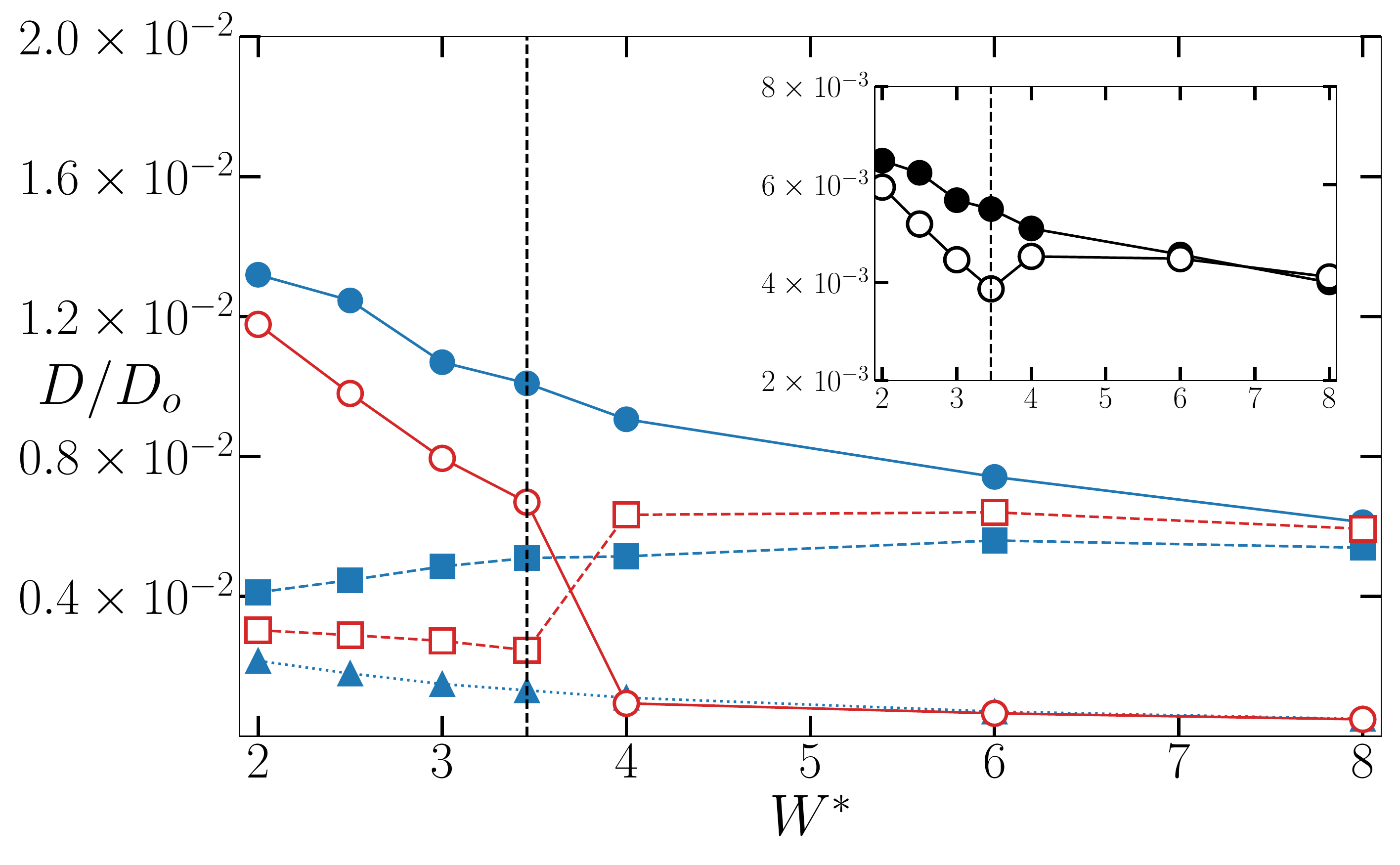}
    \caption{Self-diffusion coefficients at $\eta=0.340$, as a function of particle width and reduced by $D_{0}$. Empty circles and squares correspond, respectively, to self-diffusion coefficients calculated in the $\rm N_U$ phase along the nematic director and perpendicularly to it. Solid circles, squares and triangles refer to the self-diffusion coefficients obtained in the the $\rm N^U_B$ phase along the nematic directors $\hat{\bf n}$, $\hat{\bf m}$ and $\hat{\bf p}$, respectively. The inset reports the total (isotropic) self-diffusion coefficients for uniaxial (empty circles) and biaxial (solid circles) phases. Vertical dashed lines at $W^{*} = \sqrt{L^{*}} \approx 3.46$ indicates the transition from prolate to oblate particle shapes. Solid and dotted lines are guides for the eye.}\label{fig:DIFF1} 
\end{figure}

To better assess the dynamics of HBPs, we have calculated the self-diffusion coefficients from the slope of the MSDs at sufficiently long time scales, where $\left<\Delta r^2\right>$ changes linearly with time. The complete set of directional self-diffusion coefficients obtained in $\rm N_U^+$ and $\rm N_B$ phases for $2 \le W^* \le 8$ are shown in Fig.\,\ref{fig:DIFF1}, while the total self-diffusion coefficients are reported in the inset, both reduced by $D_0 \equiv T^{2}\tau^{-1}$. Prolate HBPs ($W^*<3.46$) exhibit an increase of their diffusion coefficient in the direction of $\hat{\bf{n}}$ upon transition from the $\rm N_U$ to the $\rm N_B^U$ phase (empty \textit{vs} solid circles). As far as the diffusion in planes perpendicular to $\hat{\bf{n}}$ is concerned, we note that the field sparks the alignment of the particle minor axes along the directors $\hat{\bf{m}}$ and $\hat{\bf{p}}$, which is not observed in $\rm N_U^+$ phase. As such, it makes sense to calculate only one self-diffusion coefficient perpendicular to $\hat{\bf{n}}$ in the $\rm N_U^+$ phase, but two distinct self-diffusion coefficients, along $\hat{\bf{m}}$ and $\hat{\bf{p}}$, in the $\rm N_B^U$ phase. Interestingly, applying an external field induces a faster dynamics along the field direction ($\hat{\bf{m}}$), but slows down the dynamics in the direction perpendicular to it ($\hat{\bf{p}}$), thus breaking the symmetry of in-plane diffusion that is observed in uniaxial nematics. We believe that these contrasting effects are due to an equilibrium between the preferential paths that a full orientational (biaxial) ordering creates and the resulting resistance to rotation that limits the ability of particles to diffuse through dense phases.

Similar considerations are also valid for oblate HBPs ($W^*>3.46$). In this case, the self-diffusion coefficient in the direction of the main nematic director $\hat{\bf{p}}$ does not change upon application of the external field (empty circles \textit{vs} solid triangles). The difference between the self-diffusivities calculated in planes perpendicular to $\hat{\bf{p}}$ exhibit similar tendencies to those reported for prolate particles, but tend to become negligible at sufficiently large particle width. In particular, at $W^*=8$, the two perpendicular self-diffusion coefficients of the $\rm N_B^U$ phase have almost the same value, which is indistinguishable from that of the parental $\rm N_U^-$ phase. A possible explanation for this behaviour is that the differences of field-induced biaxial nematics with the $\rm N_U^-$ phase are not as relevant as those with the $\rm N_U^+$ phase. More specifically, in the $\rm N_U^-$ phase, diffusion in planes perpendicular to the nematic director is enhanced by the formation of two-dimensional channels that depend on particle geometry and orientation \cite{cuetos2020, morillo2019}. In the $\rm N_B^U$ phase, these channels favour particle diffusion especially along $\hat{\bf n}$, the direction perpendicular to the surface area $WT$ that offers a lower resistance to flow than the surface area $LT$. However, the larger $W$, the less relevant this difference as the values of the self-diffusivities at $W^*=8$ confirm. In the limit of $W=L$, the two perpendicular self-diffusion coefficients should be the same. We expect a similar behaviour in systems of prolate HPBs with $W=T$ (not shown here). 

We also analyse, in the inset of Fig.\,\ref{fig:DIFF1}, the total diffusion coefficients in the $\rm N_U$ (empty symbols) and $\rm N_B^U$ (solid symbols) phases. They show a monotonic decrease with the particle width in the biaxial phase, but a more intriguing behaviour, with a minimum at the self-dual shape, in the uniaxial phase. We believe that this result, which had been also observed in recent simulations \cite{cuetos2020}, is due to dimensionality of the above-mentioned channels, being 1 in $\rm N_U^+$ phases and 2 in $\rm N_U^-$ phases. Surprisingly, at sufficiently large particle width, the difference between the total self-diffusion coefficients measured in the field-free uniaxial and field-induced biaxial phases become negligible, suggesting very similar diffusive dynamics. While the application of an external field to $\rm N_U^+$ phases produces a drastic change in structural ordering and dynamics, the same field applied to $\rm N_U^-$ phases has an effect on structure only, but it does not seem to affect dynamics. This is again due to the presence of preferential paths for diffusion: their dimensionality remains unchanged upon the field-induced uniaxial-to-biaxial transition of oblate HBPs, but increases from 1 to 2 in case the same transition is produced in systems of prolate HBPs. 

\begin{figure}[h!]
\includegraphics[width=1.00 \columnwidth]{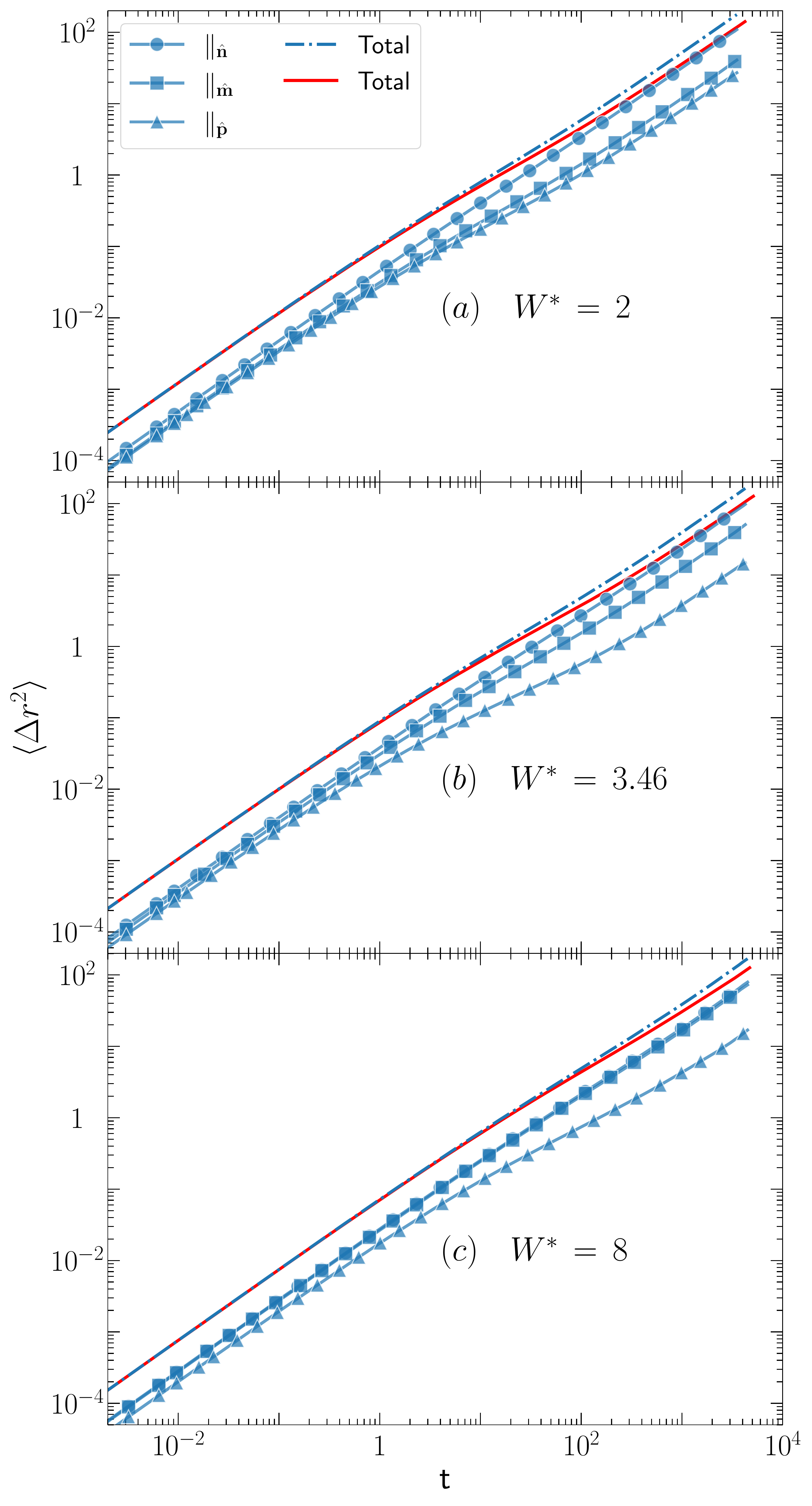}
\caption{MSD of HBPs in $\rm I$ ($\epsilon_{f}^{*} = 0$) and field-induced $\rm N_B^I$ ($\epsilon_{f}^{*} = 2$) phases at (a) $\eta=0.252$, (b) $\eta=0.307$ and (c) $\eta=0.220$. Red solid line and blue dotted-dashed line refer to the total MSDs in the I and $\rm N_{B}^{I}$ phase, respectively. Solid circles, squares and triangles refer to the MSDs obtained in the $\rm N_{B}^{I}$ phase along the nematic directors $\hat{\bf n}$, $\hat{\bf m}$ and $\hat{\bf p}$, respectively.}\label{fig:MSDISO} 
\end{figure}

In light of these observations, we now discuss the case when the same external field is applied to thermodynamically stable I phases and induces an I-to-$\rm N_B^I$ phase transition. The MSDs along the direction of the nematic directors $\bf{\hat{n}}$, $\bf{\hat{m}}$ and $\bf{\hat{p}}$ in the $\rm N_B^I$ phase are shown in Fig.\,\ref{fig:MSDISO} for prolate ($W^{*}=2$), self-dual shaped ($W^{*}=3.46$) and oblate ($W^{*}=8$) HBPs at $\eta=0.252$, $0.307$ and $0.220$, respectively. A study at the same packing fraction in the isotropic phase, as in the nematic phase, is only possible at very low values of $\eta$ (see phase diagram in Ref.\,\cite{cuetos2017}). Therefore, we have chosen to take, for each value of $W^{*}$, packing fractions close to the isotropic to nematic transition. The total MSDs, both in the field-induced $\rm N_B^I$ and parental $\rm I$ phases, are also shown for comparison. At the three packing fractions, one can observe an increase of the long-time mobility in the biaxial phase as compared to the I phase. Similarly to the increase in the long-time mobility sparked by the $\rm N_U$-to-$\rm N_B^U$ transition, also in this case the onset of two-dimensional channels boost particle diffusion with an increase in the total MSD at long time scales and for the three particle geometries studied. The directional components of the MSD in the $\rm N_B^I$ phase (along the main nematic director, along the external field and perpendicular to both) unveil a dependence on particle size that confirms the observations discussed for the $\rm N_B^u$ phase. In particular, the largest and smallest long-time MSDs are obtained, respectively, in the direction of the particle length, that is along $\hat{\bf n}$, and in the direction of $\hat{\bf{p}}$. 

\begin{figure}[h!]
    \includegraphics[width=1.00 \columnwidth]{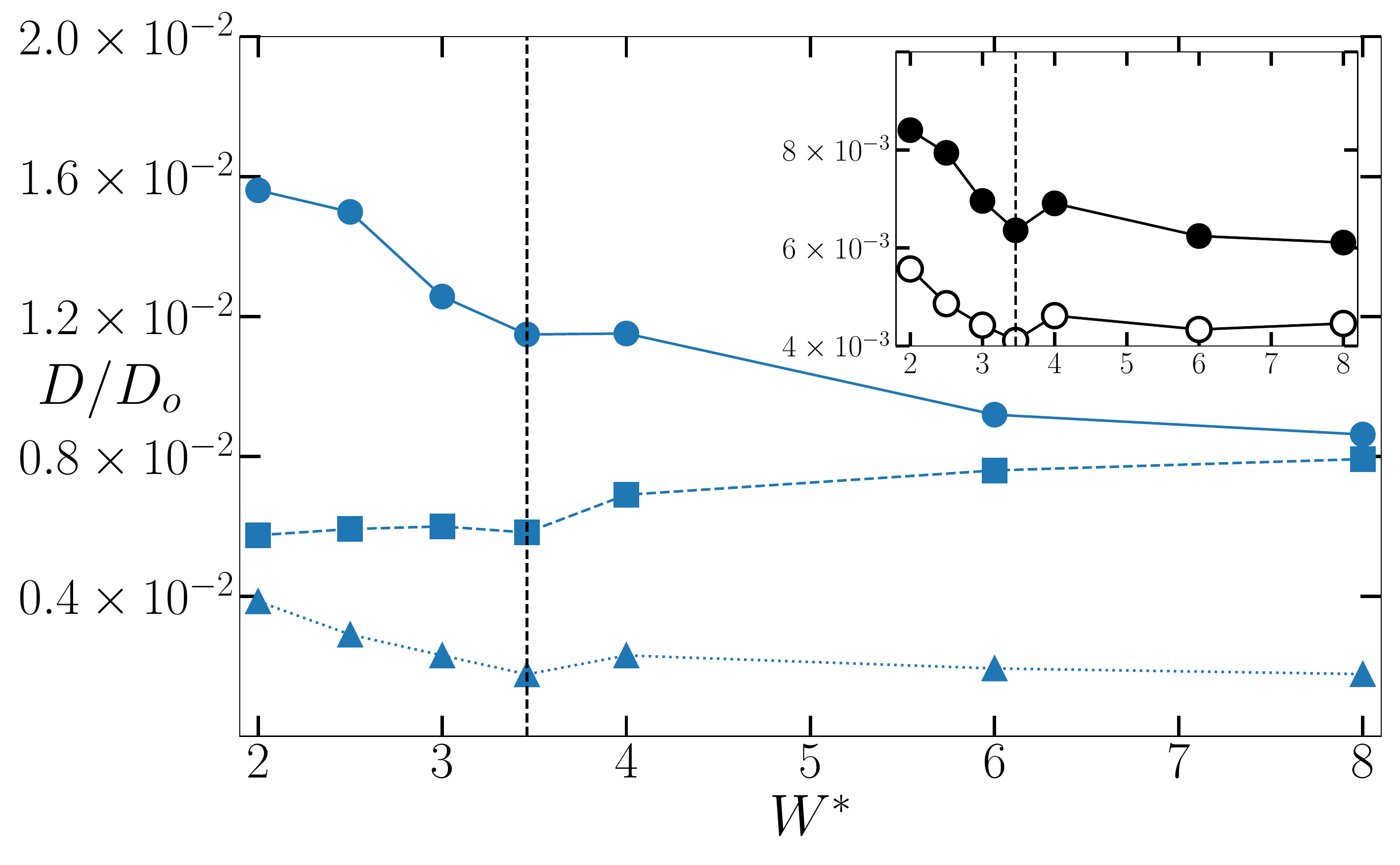}
    \caption{Self-diffusion coefficients reduced by $D_{0}$, in the induced biaxial phase $\rm N_B^I$ at packing fraction between $0.220$ and $0.307$ (see text). Solid circles, squares and triangles refer to the self-diffusion coefficients obtained along the nematic directors $\hat{\bf n}$, $\hat{\bf m}$ and $\hat{\bf p}$, respectively. Empty and solid circles in the inset refer, respectively, to the total diffusion coefficients in the parental I and the field-induced $\rm N_{B}^{I}$ phases.}\label{fig:DIFF2}
    
\end{figure}

In Fig.\,\ref{fig:DIFF2}, we show the diffusion coefficients obtained in $\rm I$ and $\rm N_B^I$ phases. One can observe that, at the prolate limit, the diffusion coefficient in the direction parallel to the main nematic director $\hat{\bf n}$ (solid circles) is larger than that in the directions perpendicular to it, while an opposite tendency is detected at the oblate limit, where the main nematic director $\hat{\bf p}$ is aligned with the particle thickness (solid triangle). This behaviour resembles that reported on the diffusion of uniaxial nematics of cuboidal \cite{cuetos2020} and spherocylindrical  particles \cite{morillo2019}, and confirms the tendencies we have discussed for $\rm N_B^U$ fluids. The analogies observed between the field-induced $\rm N_B^U$ and $\rm N_B^I$ phases suggest that, despite the differences in their orientational ordering and packing, these two phases are dynamically equivalent. Fig.\,\ref{fig:DIFF2} indicates that prolate, self-dual-shaped and oblate cuboids in $\rm N_B^I$ fluids exhibit a larger self-diffusivity along $\hat{\bf n}$ over the whole range of particle anisotropies. This self-diffusivity decreases upon increasing $W^*$ and eventually matches that along the direction $\hat{\bf m}$ of the external field at $W^*=8$. The mobility along the third nematic director, $\hat{\bf p}$, is the slowest one and does not change significantly, with a slight minimum at the self-dual shape, across the whole range of particle anisotropies. The total self-diffusion coefficients in the parental $\rm I$ and field-induced $\rm N_B^I$ phases are presented in the inset of Fig.\,\ref{fig:DIFF2}. The diffusion in the biaxial phase is significantly faster than that in the isotropic phase, but, interestingly, the qualitative behaviour is very similar, with a minimum observed at the self-dual shape in both cases. In practice, inducing a biaxial ordering leads to a faster diffusion. Moreover, by comparing the insets of Figs.\,\ref{fig:DIFF1} and \ref{fig:DIFF2}, one can observe that inducing biaxiality from I phases leads to a faster diffusion as compared to biaxial nematics induced from uniaxial phases. This difference is just a consequence of the fact that $\rm N_U$ phases ($\eta = 0.340$) are denser than I phases ($0.220  \le \eta \le 0.307$).

\begin{figure*}[t]
    \centering
   \includegraphics[width=1.00 \columnwidth]{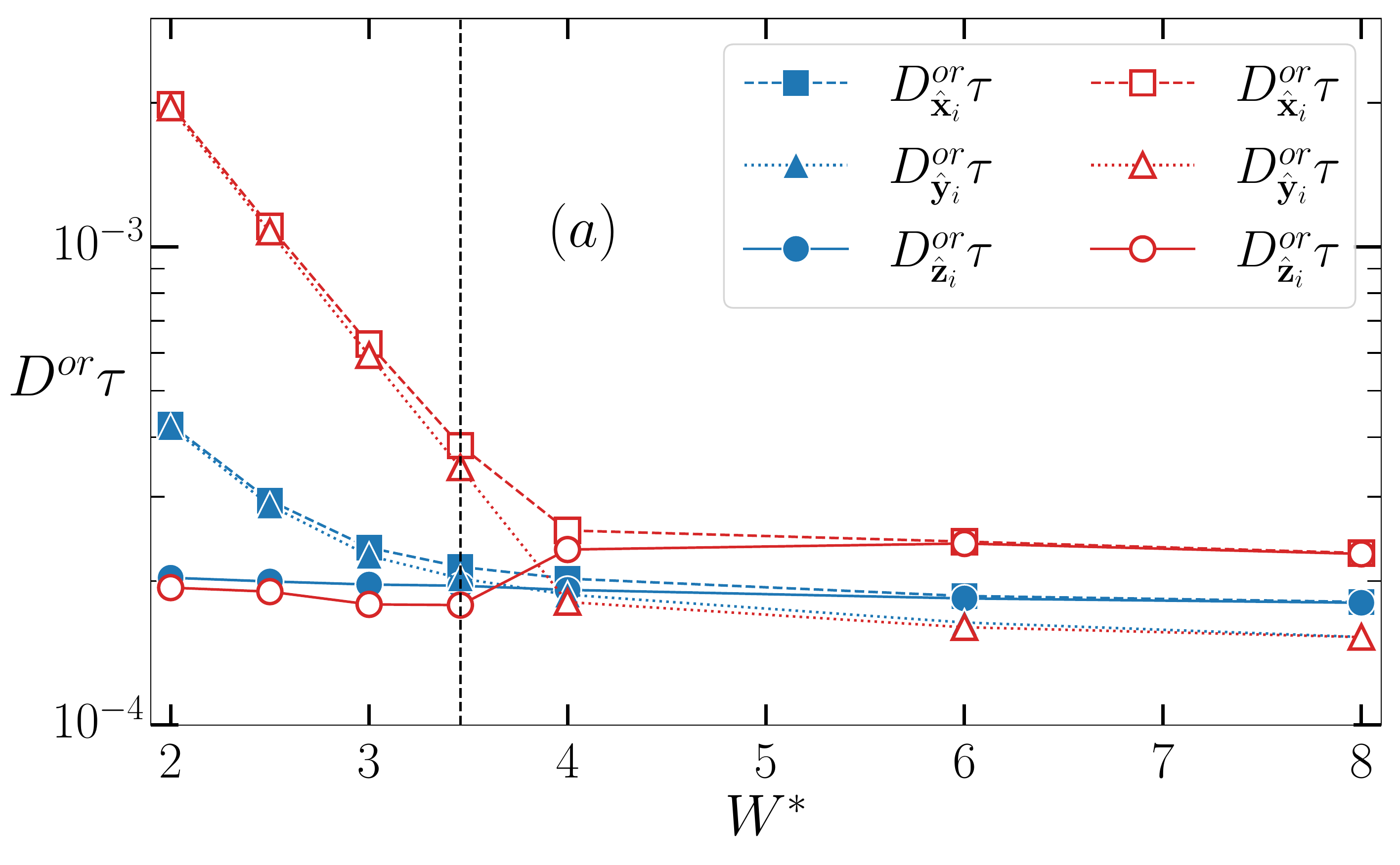}
   \includegraphics[width=1.00 \columnwidth]{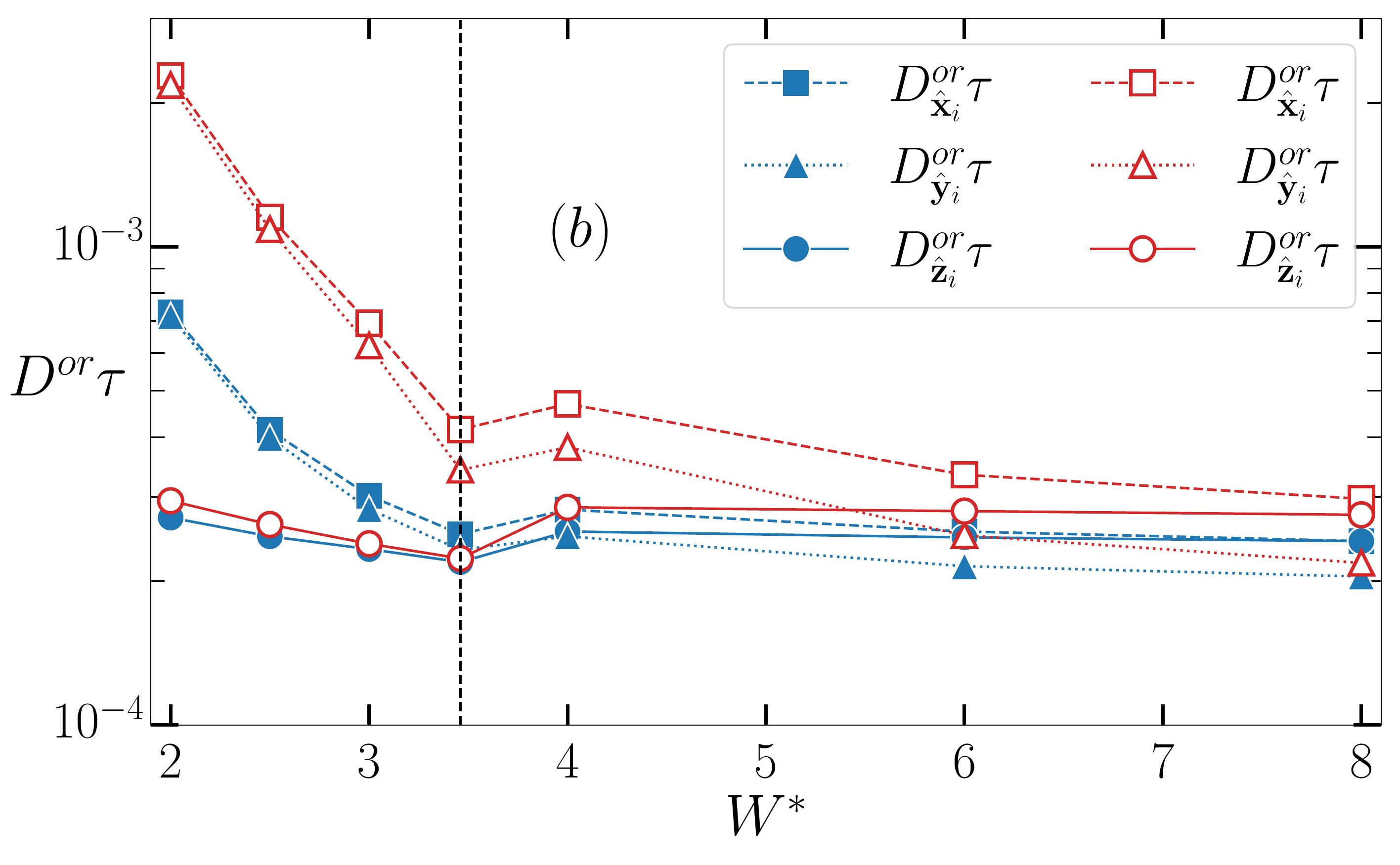}    
    \caption{Orientational self-diffusion coefficients corresponding to the rotation of the particle axes $\hat{\bf{z}}_i$ (circles), $\hat{\bf{x}}_i$ (squares), and $\hat{\bf{y}}_i$ (triangles). (a) Results obtained for $\rm N_{U}$ and $\rm N_{B}^{U}$ phases at ($\eta=0.340$). (b) Results obtained for $\rm I$ and $\rm N_{B}^{I}$ phases cases with the same range of packing fraction. The empty symbols correspond to the phases developed with external field off.}\label{fig:DIFF_OR} 
\end{figure*}

Finally, in Fig.\,\ref{fig:DIFF_OR}, we report the orientational self-diffusion coefficients of the particle unit vectors $\bf{\hat{x}}$, $\bf{\hat{y}}$ and $\bf{\hat{z}}$, respectively associated to $W$, $T$ and $L$, as calculated with Eqs.\,(\ref{eq:c}) and (\ref{eq:tau}). As a general tendency, we observe that prolate HBPs rotate faster than oblate HBPs in isotropic and nematic phases. Switching the field on enhances this difference, especially for rotations of the minor axes $\hat{\bf{x}}_i$ and $\hat{\bf{y}}_i$ around the particle length. By contrast, phase transitions have a weaker impact on the ability to rotate of oblate HBPs as can be especially appreciated in the right frame of Fig.\,\ref{fig:DIFF_OR}, reporting rotational self-diffusion coefficients in the I and $\rm N_B^I$ phases. 

%%%%%%%%%%%%%%%%%%%%%%%%%%%%%%%%%%%%%%%%%%%%
%%%%%%%%%%%%%%%%%%%%%%%%%%%%%%%%%%%%%%%%%%%%

\section{\label{sec:conclusions}Conclusions}

In summary, we have investigated the dynamics of field-induced biaxial nematics and compared it to the dynamics observed in the parental isotropic and uniaxial nematic phases. The $\rm N_B$ phase has been induced by coupling the particle intermediate axis to an external field that forces particle alignment and produces biaxiality. We stress that colloidal suspensions of cuboids are unable to spontaneously assemble into biaxial phases, unless (\textit{i}) a degree of size dispersity is incorporated \cite{effran2020}, (\textit{ii}) their aspect ratio is significantly large \cite{dussi2018}, or (\textit{iii}) an external stimulus is applied \cite{effran2021}. If none of these conditions are met, then HBPs preferentially form uniaxial nematic or smectic LCs with no evidence of biaxial nematics. In particular, investigating the response of colloidal HBPs to external fields is crucial to better understand their potential use in practical applications, especially because these stimuli are able to sensibly enrich their phase behaviour, introducing phases that cannot be observed otherwise, and their dynamics, directly modifying the particle ability to translate and rotate and hence making them more or less appealing for specific formulations. From this point of view, the rules governing the dynamics of these systems are as relevant as those regulating their phase behaviour. To this end, we have applied dynamic Monte Carlo simulation, a stochastic technique that can qualitatively and quantitatively reproduce the Brownian motion of colloids. More specifically, we have calculated the translational and rotational self-diffusion coefficients of prolate, self-dual-shaped and oblate HBPs in the uniaxial parental I and $\rm N_U$ phases as well as in the biaxial  field-induced $\rm N_B^I$ and $\rm N_B^U$ phases.

The formation of the biaxial nematic phase has an impact on the dynamical properties of prolate HBPs, but less on the dynamics of oblate HBPs. In particular, we observed that for $W \leq \sqrt{L}$, the formation of the $\rm N_B^U$ phase leads to an increase in the total self-diffusion coefficient. For this geometry, the uniaxial-to-biaxial phase transition is accompanied by an increase in the dimensionality of preferential channels for diffusion that result from the alignment of particles. Basically, the dimensionality of channels increases from 1 in field-off uniaxial phase to 3 in the field-on biaxial phase, thus enhancing the ability of HBPs to diffuse. By contrast, no change in these channels' dimensionality is observed in systems of oblate HBPs. This explains why the difference between the HBPs' dynamics in $\rm N_U^-$ and $\rm N_B^U$ phases is less relevant. For similar reasons, the most relevant differences are detected upon transition from the I phase, which does not present preferential channels, to the $\rm N_B^I$ phase, whose channels are observed along the three nematic directors. Remarkably, for a given particle width, the diffusion coefficient in the $N_B^I$ phase is larger than in the $\rm N^U_B$ phase. Although in this case the nematic order is higher, the lower packing in the $\rm N_B^I$ phases seems to play a relevant role. This is important if biaxial materials with short response times are to be designed. Therefore, the higher orientational order in the biaxial phase results in an increase of the diffusion coefficients overall and in the direction of the applied field. Consequently, the diffusion channels have a considerable impact on the orientational self-diffusion coefficients, with a generalized decrease in values; except in the case of vector $\bf{\hat{z}}$ for prolate cuboids in the $\rm N^U_B$ phase, where it increases slightly, and also in the case of vector $\bf{\hat{y}}$ for oblate cuboids in the same phase, where changes of the same value are not observed with respect to the case without field. Comparing this last case with the results obtained at lower packing fraction, we observe a decrease in the orientational self-diffusion coefficient when applying the field, which indicates the formation of the channels and the consequent increase in the translational diffusion coefficients for $W > \sqrt{L}$ cases. Conversely, a higher correlation of the $\bf{\hat{y}}$ vectors results in a lower diffusion along the corresponding director. This is similar to what happens in $\rm N_U^-$ phases \cite{cuetos2020}, and is a consequence of steric hindrances for particles to diffuse in this direction.

%%%%%%%%%%%%%%%%%%%%%%%%%%%%%%%%%%%%%%%%%%%%

\begin{acknowledgments}
A.C. and A.R.-R. acknowledge the Consejería de Transformación Económica, Industria, Conocimiento y Universidades de la Junta de Andalucía/FEDER for funding through project P20-00816. A.C. also acknowledge funding from the Spanish Ministerio de Ciencia, Innovación y Universidades and FEDER (Project no. PGC2018-097151-B-I00). A.R.-R. also acknowledges financial support from Consejería de Transformación Económica, Industria, Conocimiento y Universidades de la Junta de Andaluc\'ia through post-doctoral grant no. DC 00316 (PAIDI 2020), co-funded by the EU Fondo Social Europeo (FSE). A.P. is supported by a “Maria Zambrano Senior” distinguished researcher fellowship, financed by the European Union within the NextGenerationEU program. We thank C3UPO for the HPC facilities provided.
\end{acknowledgments}

%\appendix

%\section{Appendixes}
%\verb+\appendix*+ command

%\section{A little more on appendixes}

\nocite{*}
\bibliography{biaxfield}% Produces the bibliography via BibTeX.

\end{document}